\documentclass[
    ,final            
  ]
  {aipproc}

\layoutstyle{6x9}

\newcommand{\pipi}{\pi^{+}\pi^{-}}
\newcommand{\KK}{K^{+}K^{-}}
\newcommand{\ppbar}{p\overline{p}}

\newcommand{\ee}{e^{+}e^{-}}

\newcommand{\eetomm}{\ee \rightarrow m^{+}m^{-}}

\newcommand{\mff}{|F_{m}(s)|}

\begin{document}

\title{Precision Measurements of the Charged Pion, Charged Kaon, 
and Proton Electromagnetic Form Factors at $s$ = 13.48 GeV$^2$}


\author{Peter Zweber (for the CLEO Collaboration)}{
  address={Northwestern University, Evanston, Illinois, 60208, USA}
}

\begin{abstract}
Using 20.7 pb$^{-1}$ of $e^+e^-$ annihilation data taken at $\sqrt{s}=3.671$ GeV 
with the CLEO--c detector, precision measurements of the electromagnetic form 
factors of the charged pion, charged kaon, and proton have been made for timelike 
momentum transfer of $|Q^2|=13.48$ GeV$^2$ by the reaction $e^+e^-\to h^+h^-$.  
The measurements are the first ever with identified pions and kaons of $|Q^2|>4$ 
GeV$^2$, with the results 
$|F_\pi(13.48\;\mathrm{GeV}^2)|=0.075\pm0.008(\mathrm{stat})\pm0.005(\mathrm{syst})$ 
and 
$|F_K(13.48\;\mathrm{GeV}^2)|=0.063\pm0.004(\mathrm{stat})\pm0.001(\mathrm{syst})$.  
The result for the proton, assuming $|G^p_E|=|G^p_M|$, is 
$|G^p_M(13.48\;\mathrm{GeV}^2)|=0.014\pm0.002(\mathrm{stat})\pm0.001(\mathrm{syst})$, 
which is in agreement with earlier results. 
\end{abstract}

\maketitle


Electromagnetic form factors of hadrons are among the most important physical 
observables.  They provide direct insight into the distribution of charges, 
currents, color, and flavor in the hadron.  We report on the first precision 
measurements for the timelike electromagnetic form factors of the 
pion, kaon, and proton, for $s=-Q^2=13.48$ GeV$^2$, by means of the reactions
\begin{equation}
e^+e^-\to\pi^+\pi^-,\;K^+K^-,\;\mathrm{and}\;p\bar{p}
\end{equation}

Theoretical predictions of form factors based on perturbative Quantum Chromodynamics 
(pQCD) rely on the validity of 
factorization for sufficiently high momentum transfers, and lead to quark 
counting rules, which predict \cite{quarkcounting} that 
$|F(Q^2)|\propto |Q^2|^{1-n}$, where $n$ is the number of quarks, so that 
$|F(Q^2)|\propto \alpha_S/|Q^2|$ for mesons and 
$|F(Q^2)|\propto \alpha_S^2/|Q^4|$ for baryons. pQCD also predicts \cite{pQCD} 
that the form factors for the helicity--zero mesons $m=\pi,\;K,\;\rho,\;...$ are 
proportional to the squares of their decay constants so that 
$|F_\pi(Q^2)|/|F_K(Q^2)|=f_\pi^2/f_K^2$, as $|Q^2|\to\infty$.

The timelike form factors of the charged helicity--zero mesons, $\mff$, are related 
to the differential cross section for their pair production by
\begin{equation}
\frac{d\sigma_{0}(s)}{d\Omega}(\eetomm) = \frac{\alpha^{2}}{8s}~
\beta^{3}_{m}~\mff^{2}\mathrm{sin}^{2}\theta,
\label{mesonad}
\end{equation}
where $m$ = $\pi$ or $K$, $\alpha$ is the fine-structure constant, 
$\beta_{m}$ is the meson velocity in the laboratory system, 
$s$ is the center-of-mass energy squared, and 
$\theta$ is the laboratory angle between the meson and the positron beam.  

The $e^+e^-\to p\bar{p}$ differential cross section is related to the magnetic 
($G^p_M$) and electric ($G^p_E$) form factors of the proton. With $\tau=4m_p^2/s$
\begin{equation}
\frac{d\sigma_0(s)}{d\Omega} \! =\! \frac{\alpha^2}{4s} \beta_p \!\!\left[ 
|G_M^p(s)|^2(1\! +\!\cos^2\!\theta)\! +\! \tau |G^p_E(s)|^2\sin^2\!\theta\right]
\label{baryonad}
\end{equation}
In the present measurements we do not have sufficient statistics to separate 
$|G^p_E(s)|$ and $|G^p_M(s)|$.  We therefore analyze our data with two 
assumptions, $|G^p_E| = |G^p_M|$ and $|G^p_E| = 0$. 

The $e^+e^-$ annihilation data sample used in the present measurements consists 
of 20.7 pb$^{-1}$ taken at $\sqrt{s}=3.671$ GeV. 
The data were collected at the Cornell Electron 
Storage Ring (CESR) with the detector in the CLEO--c 
configuration.  Details of the detector performance, event selection, and 
analysis procedure are described in Ref. \cite{FFPaper}.

\begin{figure}
  \includegraphics[height=0.24\textheight]{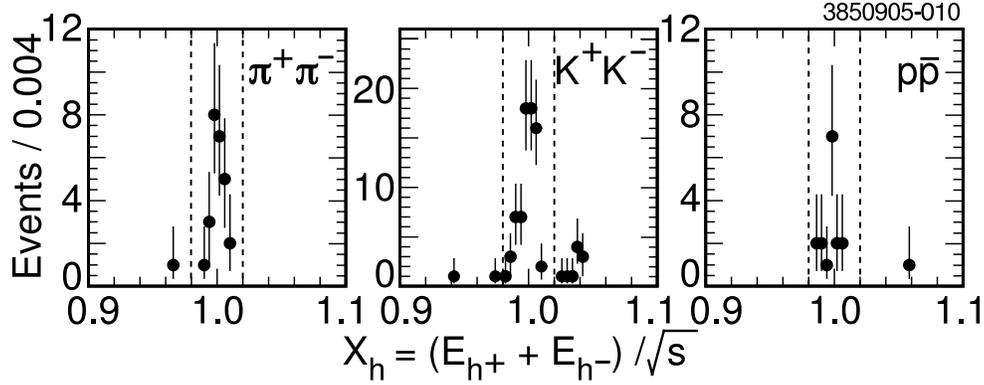}
  \caption{Data events as a function of $X_{h}$ for $\pipi$ (left), $\KK$ (middle), 
  and $\ppbar$ (right) final states.  The dashed lines denote the signal regions 
  defined as 0.98 $<$ $X_{h}$ $<$ 1.02.}
  \label{fig:xwidedata}
\end{figure}

Figure \ref{fig:xwidedata} shows the distribution of the selected 
events as a function $X_h = (E_{h^{+}}+E_{h^{-}})/\sqrt{s}$.  
The signal region is defined as $0.98<X_h<1.02$ as bounded by the dashed lines.  
The resulting cross sections are listed in Table 1.  
Various sources of systematic uncertainties in the cross sections have been studied.  
Their sum in quadrature is 14.6\% for pions, 4.4\% for 
kaons, and 8.9\% for protons.  Integrating Eqs. \ref{mesonad} and 
\ref{baryonad} leads us to our final results 
for the form factors as listed in Table 1.

\begin{figure}[!tb]
\begin{tabular}{ll}
\includegraphics[height=0.36\textheight]{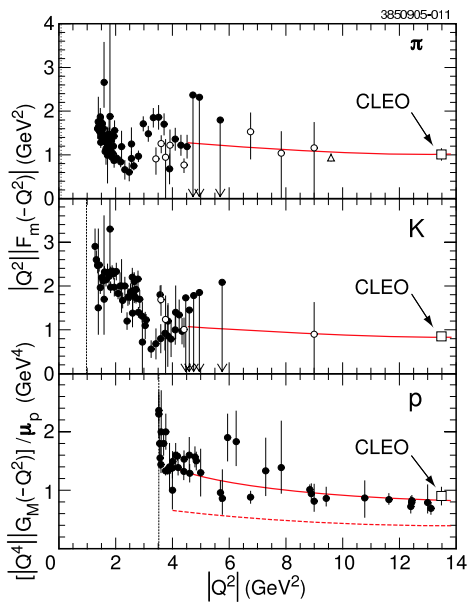}
&
\raisebox{1.6in}{\parbox{3.5in}{
{\small \textbf{TABLE 1.}~~Summary of form factor results.  
The first errors are statistical and the second errors are systematic.  
The form factor results for protons correspond to the assumption $|G^p_E| = |G^p_M|$.  
Results for the assumption $|G^p_E| = 0$ are $\sim$9$\%$ larger.}\\
\begin{footnotesize}
\begin{tabular}{lcc}
\hline
$h$ & $\sigma(e^+e^-\to h^+h^-)$ (pb) & $|F(13.48\;\mathrm{GeV}^2)|$ \\
\hline
$\pi$ & $9.0\pm1.8\pm1.3$ & $|F_\pi|=0.075\pm0.008\pm0.005$ \\
$K$   & $5.7\pm0.7\pm0.3$ & $|F_K|=0.063\pm0.004\pm0.001$ \\
$p$   & $1.2\pm0.4\pm0.1$ & $|G^p_M|=0.014\pm0.002\pm0.001$ \\
\hline
\end{tabular}
\end{footnotesize}}}
\\
\end{tabular}
\caption{Open squares show the present results for the 
pion (top), kaon (middle), and proton (bottom) form factors.
Other symbols show a compilation of results in the literature.  
For details, see Refs. \cite{timelikepionkaon, e760e835}.  
The arbitrarily normalized solid curves show the variation of $\alpha_S(|Q^2|)$ 
(top and middle) and $\alpha_S^2(|Q^2|)$ (bottom).  
The dashed curve in the bottom plot shows the $\alpha_S^2(|Q^2|)$ 
fit to the spacelike form factors of the proton.}
\label{fig:tlffcomb} 
\end{figure}

Our results for timelike form factors are displayed in 
Figure \ref{fig:tlffcomb} as $|Q^2||F_\pi|$, $|Q^2||F_K|$, 
and $|Q^4||G_M^p|/\mu_p$, together with the existing world data 
\cite{timelikepionkaon,e760e835} for the same. 

Our precision results for the charged pion and kaon stand alone at present.  
Our results for the pion form factor provides empirical validity for 
$|Q^2||F_\pi(9.6~\mathrm{GeV}^2)|$ = (0.94$\pm$0.06) GeV$^2$ obtained by interpreting 
$\Gamma(J/\psi\to\pi^+\pi^-)/\Gamma(J/\psi\to e^+e^-)$ as a measure of the pion 
form factor \cite{jpsiff}.   Together, the two appear to support the pQCD 
prediction of $\alpha_S/|Q^2|$ variation of the form factor at large $|Q^2|$.  
The pQCD prediction $|F_\pi(Q^2)|/|F_K(Q^2)|$ 
= $f_\pi^2/f_K^2$ = 0.67$\pm$0.01 \cite{pdg} is in disagreement with our result 
$|F_\pi(13.48\;\mathrm{GeV}^2)|/|F_K(13.48~\mathrm{GeV}^2)|$ = 1.19$\pm$0.17.  
Bebek {\sl et al.}~\cite{bebek} have reported $|Q^2||F_\pi(9.77~\mathrm{GeV}^2)|$ = 
0.69$\pm$0.19 GeV$^2$ for the spacelike form factor.  Within errors this is 
consistent with being nearly factor two smaller than the timelike form factors 
for $|Q^2| > 9$ GeV$^2$, as found for protons.  

Our result for $|G^p_M(13.48~\mathrm{GeV}^2)|$ provides the 
first independent confirmation of the Fermilab E760/E835 experiments, which 
measured the reverse reaction $p\bar{p}\to e^+e^-$ \cite{e760e835}.


\begin{theacknowledgments}
We gratefully acknowledge the effort of the CESR staff in providing us with 
excellent luminosity and running conditions. This work was supported by the 
National Science Foundation and the U.S. Department of Energy.
\end{theacknowledgments}



\begin{thebibliography}{99}

\bibitem{quarkcounting} S. J. Brodsky and G. R. Farrar, 
Phys. Rev. Lett. \textbf{31}, 1153 (1973); 
Phys. Rev. \textbf{D 11}, 1309 (1975).

\bibitem{pQCD} G. R. Farrar and D. R. Jackson, 
Phys. Rev. Lett. \textbf{43}, 246 (1979); 
S. J. Brodsky and G. P. Lepage, Phys. Lett. \textbf{B 87}, 359 (1979); 
V. Efremov and A. V. Radyushkin, Phys. Lett. \textbf{B 94}, 245 (1980); 
G. P. Lepage and S. J. Brodsky, Phys. Rev. \textbf{D 22}, 2157 (1980).

\bibitem{FFPaper}T. K. Pedlar \textit{et al.} (CLEO Collaboration), 
Phys. Rev. Lett. \textbf{95}, 261803 (2005).

\bibitem{timelikepionkaon} For a compilation of pion and kaon form factor data, 
see M. R. Whalley, J. Phys. \textbf{G 29}, A1 (2003).

\bibitem{e760e835} M. Andreotti {\sl et al.} (E835 Collaboration), 
Phys. Lett. \textbf{B 559}, 20 (2003), and references therein.

\bibitem{jpsiff} R. Kahler and J. Milana, Phys. Rev. \textbf{D 47}, R3690 (1993); 
J. Milana, S. Nussinov, and M. G. Olsson, Phys. Rev. Lett. \textbf{71}, 2533 (1993).

\bibitem{pdg} Review of Particle Properties, 
S. Eidelman {\itshape{et al.}}, Phys. Lett. {\bf B 592} , 1 (2004).

\bibitem{bebek} C. J. Bebek \textit{et al.}, 
Phys. Rev. \textbf{D 17}, 1693 (1978). 

\end{thebibliography}
\end{document}